\documentclass[twocolumn]{aastex701}%linenumbers, 

\usepackage{amsmath}

\def\Gyr{{\rm\,Gyr}}
\def\kpc{{\rm\,kpc}}

\submitjournal{Accepted on August 6, 2025 in The Astrophysical Journal Letter}

\shorttitle{Chemical Analysis of the Galaxy}
\shortauthors{Malhan}

\begin{document}

\title{Milky Way's Metal-Poor Stars display Chemical Transition near the Solar Radius}

\correspondingauthor{Khyati Malhan}
\email{kmalhan07@gmail.com}

\author[0000-0002-8318-433X]{Khyati Malhan}
\affiliation{Independent researcher}
\email{kmalhan07@gmail.com}
%\nocollaboration{1}

\begin{abstract}
The metal-poor stars of a galaxy offer insights into that galaxy's early formation processes and accretion history. Here, we investigate whether the metal-poor stars of our Milky Way galaxy exhibit any characteristic trends in Galactocentric distance versus chemical abundances -- i.e. in the space of $r_{\rm GC}$ vs. [Fe/H] and $r_{\rm GC}$ vs. [X/Fe] -- and if yes, then what is their implication for Galaxy formation. We combine the datasets of APOGEE~DR17 and \textit{Gaia}~DR3, where the former provides stellar abundances and the latter provides stellar parallaxes. We analyze bright ($G<13$) and metal-poor ([Fe/H]$<-1.2$) stars located far from the disk ($|z|\geq1$~kpc), and explore a total of $19$ abundances. We find that $9$ different abundances exhibit a drastic transition in their distribution near the Solar radius $r_{\rm GC}=8$~kpc. This trend is very unlikely to be related to radial migration, as our metal-poor sample does not contain any disk star. We also analyze the Gaia-Sausage/Enceladus stars, which is a dominant metal-poor population of the Galaxy, and find that it alone cannot account for this trend. This suggests that the Milky Way's metal-poor populations inside and outside the Solar radius likely originated from distinct chemical enrichment scenarios and formation processes.
\end{abstract}
%\keywords{Milky Way formation -- Galaxy abundances -- Surveys}
%Galaxy: formation - Galaxy: abundances - surveys}

%%%%%%%%%%%%%%%%%%%%%%%%%%%%%%%%%%%%%%%%%%%%%%%%%%%%%%%%%%%%%%%%%%
%%%%%%%%%%%%%%%%% BODY OF PAPER %%%%%%%%%%%%%%%%%%%%%%%%%%%%%%%%%%
%%%%%%%%%%%%%%%%%%%%%%%%%%%%%%%%%%%%%%%%%%%%%%%%%%%%%%%%%%%%%%%%%%
\section{Introduction}\label{sec:Introduction}

A key focus of the Galactic archaeology field is to determine the Milky Way's formation processes and its accretion history \citep[see e.g.][]{2002ARA&A..40..487F, 2020ARA&A..58..205H, 2024NewAR..9901706D}. In this regard, a powerful approach is to use the Milky Way's observations and study its metal-poor stars, i.e. stars with metallicity [Fe/H]$\lesssim-1$. These stars are generally assumed to be the oldest stellar populations, with age $\gtrsim11\Gyr$, which were initially formed inside foreign, low-mass progenitors that later accreted into the Milky Way \citep[see e.g.][]{2015ARA&A..53..631F}; although recent studies suggest that at least some of the metal-poor stars were formed in-situ \citep[see e.g.][]{2022MNRAS.514..689B, 2023ApJ...953..143F, 2024MNRAS.534.1985B, 2025MNRAS.537.3730H}. Therefore, analyzing metal-poor stars and understanding their chemical abundances and distribution in the Galaxy enables us to unravel the Milky Way's early formation history and chemical enrichment scenarios \citep[see e.g.][]{2015MNRAS.453..758H, 2018ApJ...863..113H, 2019ApJ...887..237C, 2019MNRAS.489.5900D, 2020A&A...636A.115D, 2021A&A...650A.110M, 2022ApJ...930L...9M, 2022ApJ...930..103Y, 2022ApJ...928...23E, 2022MNRAS.514..689B, 2022ApJ...941...45R, 2023MNRAS.520.5671H, 2022MNRAS.515.4082A, 2022ApJ...930...47H, 2023MNRAS.525.2208R, 2023A&A...670L...2D, 2023MNRAS.518.4557S, 2024ApJ...977..278R, 2024arXiv240817250V, 2024MNRAS.533.2420M, 2024ApJ...972..112C, 2024ApJ...964..104M, 2025arXiv250307738S, 2025JApA...46...15S}.

In the present study, our aim is to examine the metallicity [Fe/H] and element abundances [X/Fe]s of the metal-poor stars as a function of the Galactocentric distance $r_{\rm GC}$, and identify potential trends in these distributions and understand their possible origin. The motivation in doing so is that: if the Galaxy formed inside-out, then those stars that currently populate the inner regions of the Galaxy would have originated earlier than those in the outer regions, regardless of whether their origin stems from accretion processes, secular processes, or a combination of both. This implies that $r_{\rm GC}$ serves as a rough proxy for the timeline of the Galaxy formation, and this makes $r_{\rm GC}$ vs. [Fe/H] and $r_{\rm GC}$ vs. [X/Fe] as potential tracers of the Galaxy's chemical evolution. While the radial migration of stars can wash out some of these potential correlations or create new ones \citep[see e.g. ][]{2002MNRAS.336..785S, 2018A&A...616A..86H, 2021ApJ...919...52Z, 2024A&A...690A.147H, 2024MNRAS.533..538L}, its effects are strongest within the Galactic disk and less significant in the Galactic halo. And the halo is primarily composed of metal-poor stars. Therefore, if the $r_{\rm GC}$ vs. [Fe/H] or $r_{\rm GC}$ vs. [X/Fe] distributions of metal-poor stars show any characteristic transitions at specific $r_{\rm GC}$ values, then this may be linked to distinct phases in Galaxy's chemical enrichment. To this end, Section~\ref{sec:Data} describes the data used, section~\ref{sec:Analysis} details the analysis and we conclude in Section~\ref{sec:Discussion}.

\begin{figure}%[h]
\begin{center}
%\vspace{-0.3cm}
\includegraphics[width=\hsize]{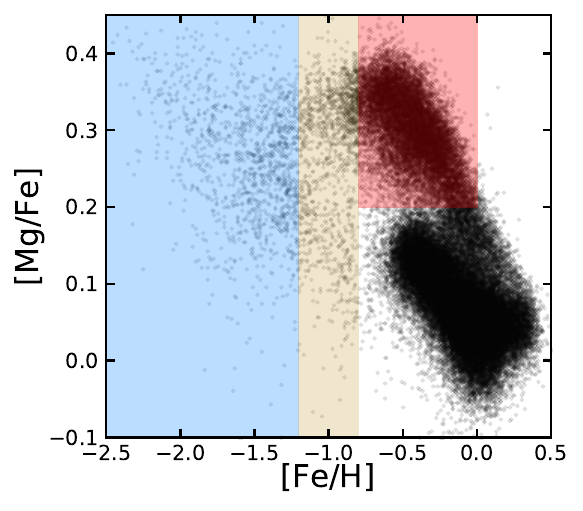}
\end{center}
\vspace{-0.5cm}
\caption{Overview of the stars in the [Fe/H] vs. [Mg/Fe] space. Different shaded regions represent different sample selections.}
\label{fig:Fig_1}
\end{figure}
\section{Data}\label{sec:Data}

For our analysis, we require such a dataset where each star possesses metallicity [Fe/H], element abundance [X/Fe]s and Galactocentric distance $r_{\rm GC}$. To achieve this, we do the following.

We cross-match APOGEE~DR17 \citep{2022ApJS..259...35A} with \textit{Gaia}~DR3 \citep{2023A&A...674A...1G}, producing a sample of $682284$ stars. The former catalog provides [Fe/H] and [X/Fe]s and the latter provides parallaxes. Next, we apply certain APOGEE-based quality cuts to ensure good abundances and to minimize contamination by dwarf stars and magellanic cloud stars; same as the implementation of \cite{2020MNRAS.492.1641M, 2023MNRAS.520.5671H, 2023A&A...676A.108C}. We retain only those stars possessing \texttt{SNR}$\gtrsim50$, \texttt{STARFLAG}$=0$, $3500 \leq T_{\rm eff} <5500$,  log(g)$<3.6$, and \texttt{PROGRAMNAME}$\neq$\texttt{magclouds}, reducing the sample to $218736$ stars. At this stage we do not use the \texttt{X\_FE\_FLAG} parameter, but implement this in Section~\ref{sec:Analysis} when we analyze the individual abundance parameters. Next, we apply certain quality cuts to ensure good \textit{Gaia} astrometry; same as the implementaion of \cite{2024ApJ...964..104M}. We retain only those stars which possess \texttt{parallax\_over\_error}$\geq 5$, as this ensures that the resulting sample contains well-measured parallaxes with relative parallax error of $\leq20\%$. Effectively, the resulting stars possess $\varpi>0$, which renders the estimated heliocentric distances $(=1/\varpi)$ physical\footnote{The median distance uncertainty of our sample is $0.025\kpc$. Since this uncertainty is small, this justifies our choice of distance calculation as $1/\varpi$.}. Next, we correct for the parallax zero point of each star as $\varpi_{\rm corrected} = \varpi_{\rm observed}-(-0.017)$ \citep[see e.g.][]{2021A&A...649A...2L}. Next, we retain only those stars with \texttt{phot\_bp\_rp\_excess\_factor}$< 1.7$, because this cut removes those stars with strong $G_{\rm BP}- G_{\rm RP}$ color excess that occurs in the very crowded regions of the Galaxy \citep[see e.g.][]{2018A&A...616A..17A}\footnote{For normal stars, \texttt{phot\_bp\_rp\_excess\_factor}$\approx1$.}. Moreover, we retain only those stars which posses \texttt{ruwe}$< 1.4$, because this value has been prescribed as a quality cut for ``good'' astrometric solutions by \cite{2021A&A...649A...2L}. These Gaia selection cuts reduce the sample to $92054$ stars. Next, we correct \textit{Gaia}'s photometry for dust extinction using \cite{1998ApJ...500..525S} maps and assuming the extinction ratios $A_{\rm G}/A_{\rm V} = 0.86117$, $A_{\rm BP}/A_{\rm V} = 1.06126$, $A_{\rm RP}/A_{\rm V} = 0.64753$ as listed on the web interface of the PARSEC isochrones \citep{2012MNRAS.427..127B}. Henceforth, all magnitudes and colors refer to these extinction-corrected values. To ensure completeness across the sky we further retain only those stars with $G_0<12.8$, reducing the sample to $65051$ stars. Next, we remove all the stars that lie within seven tidal radii of the globular clusters in the projection space, using the \cite{2010arXiv1012.3224H} catalog. The resulting sample comprises $62419$ stars and are shown in Fig.~\ref{fig:Fig_1}.

\begin{figure}
\begin{center}
%\vspace{-0.3cm}
\includegraphics[width=\hsize]{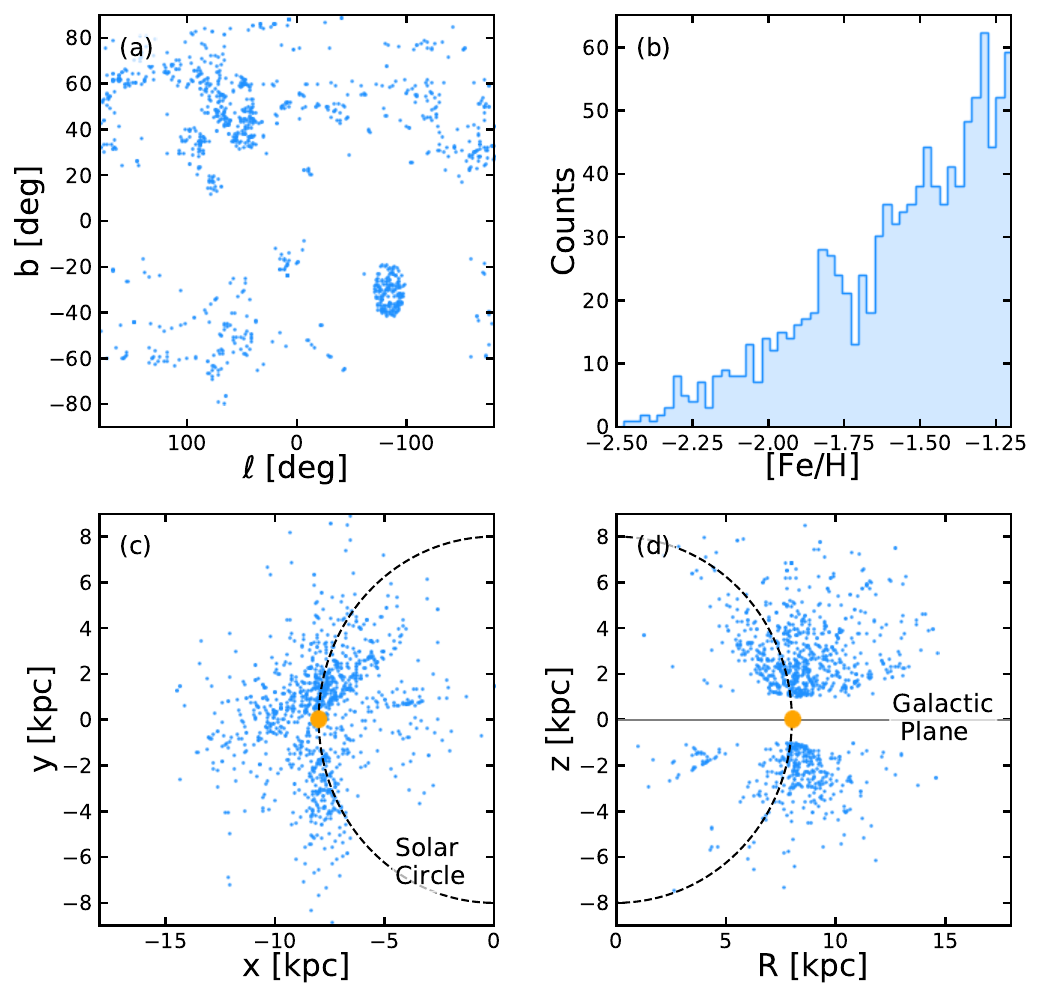}
\end{center}
\vspace{-0.5cm}
\caption{Overview of the metal-poor sample. Panel `a' shows the on-sky distribution in Galactic coordinates, panel `b' shows the metallicity distribution function, panel `c' shows the top-view of the Galaxy in the Cartesian $x-y$ plane and panel `d' shows the Galactic $R-z$ plane. The orange dot denotes the Sun.}
\label{fig:Fig_2}
\end{figure}

We create the metal-poor sample by selecting those stars with [Fe/H]$<-1.2$ and that lie farther than $|\rm{z}|\geq1\kpc$ from the Galactic plane, resulting in $1068$ stars. The $|\rm{z}|$ criteria ensures that these stars are minimally affected by radial migration, if at all. This [Fe/H] cut removes the [$\alpha$/Fe]-rich ``thick'' disk stars \citep[see e.g.][]{2019MNRAS.482.3426M} and also those stars lying in the range $-1.2\lesssim$[Fe/H]$\lesssim-1$ which may correspond to the transition regime from the thick disk to the halo \citep[see e.g.][]{2015MNRAS.453..758H, 2024ApJ...972..112C}. We also create two additional samples: the thick-disk sample (by selecting stars that follow $-0.8\leq$[Fe/H]$<0$ and $0.2\leq$[Mg/Fe]$<0.45$) and thick-disk-to-halo sample (stars that follow $-1.2\leq$[Fe/H]$<-0.8$). These two samples are required for drawing comparisons with the metal-poor sample, as discussed in Section~\ref{sec:Analysis}. These three sample selections are highlighted in Fig.~\ref{fig:Fig_1}. In Fig.~\ref{fig:Fig_2}, we provide an overview of the metal-poor sample, and the same for the other two samples are shown in the appendix Fig.~\ref{fig:Fig_appendix1}. While computing the Galactocentric cartesian distances of stars, we use the Sun's Galactocentric location as $8.178\kpc$ \citep{2019A&A...625L..10G}.

\begin{figure*}[t!]
\begin{center}
\vspace{-0.3cm}
\includegraphics[width=\hsize]{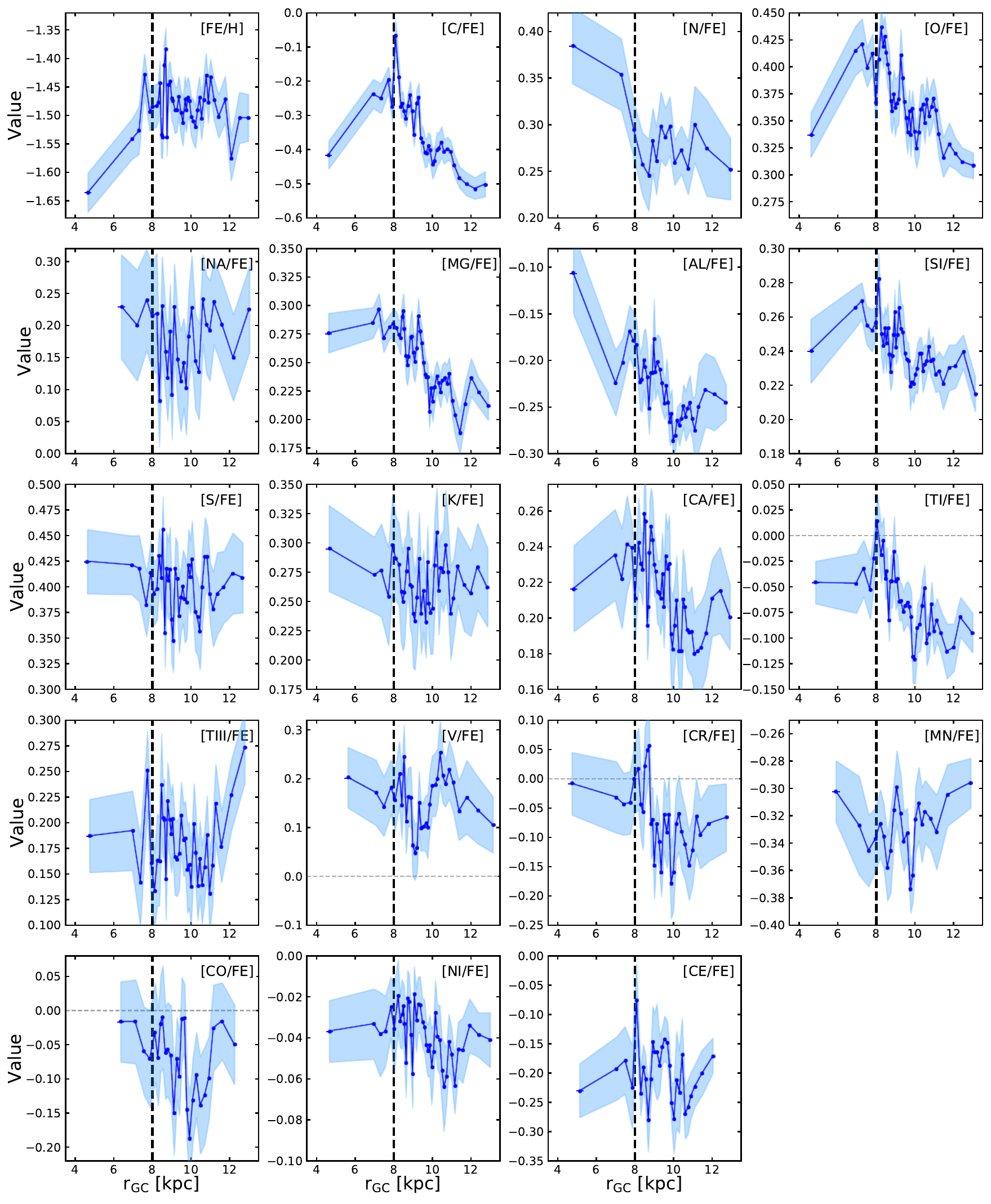}
\end{center}
\vspace{-0.5cm}
\caption{$r_{\rm GC}$ vs. [Fe/H] and $r_{\rm GC}$ vs. [X/Fe]s of the metal-poor sample. In each panel, the vertical line at $r_{\rm GC} = 8\kpc$ denotes the Solar radius. The horizontal lines (visible in some panels) are drawn where Y-axis$=0$.}
\label{fig:Fig_3}
\end{figure*}
\section{Analysis}\label{sec:Analysis}

We find that plotting simply the $r_{\rm GC}$ vs. [Fe/H] and $r_{\rm GC}$ vs. [X/Fe]s of the individual stars does not clearly reveal the underlying trend, as can be seen in the appendix Fig.~\ref{fig:Fig_appendix2}. This is because the plot is scatter dominated and this obscures the underlying trend. Therefore, we proceed with computing the median trends.

Fig.~\ref{fig:Fig_3} shows the median trends in $r_{\rm GC}$ vs. [Fe/H] and $r_{\rm GC}$ vs. [X/Fe]s distributions of the metal-poor stars. We compute these trends by accounting for the scatter and the measurement uncertainties in parallaxes, [Fe/H] and [X/Fe]. We also ensure that each $r_{\rm GC}$ bin possesses large and same number of stars so that the statistical value of each bin is robust and comparable. This is achieved as follows.

To plot the $r_{\rm GC}$ vs. [Fe/H] distribution, we do the following. We generate $10$ realizations of the metal-poor sample by sampling $r_{\rm GC}$ and [Fe/H] for each star based on their measured means and uncertainties. Particularly for $r_{\rm GC}$, we actually sample the $\varpi$ parameter and then convert it into the Galactocentric distance using the same value for the Sun's Galactocentric location as described in Section~\ref{sec:Data}. During this sampling procedure, we use only those stars that possess APOGEE parameter \texttt{FE\_H\_FLAG}$=0$ to ensure high quality abundance measurements. That is, if $n$ number of stars possess \texttt{FE\_H\_FLAG}$=0$, this sampling procedure creates a total of $n\times10$ number of sampled data values. Next, we create $r_{\rm GC}$ bins of varying sizes such that each bin contains sampled data values from exactly $75$ unique stars. %Note that the actual number of data points per bin varies a lot, from $120$ to $705$, due to large uncertainties in $\varpi$. 
This occupation number allows for enough number of $r_{\rm GC}$ bins so that the resulting trend becomes discernible, especially in the inner Galaxy region with $r_{\rm GC}<8\kpc$ where the sample size is smaller. Furthermore, this occupation number is large enough to ensure robust statistical estimate within each bin. In each bin, we compute the median values of $r_{\rm GC}$ and [Fe/H], and these values are plotted as the solid curve in Fig.~\ref{fig:Fig_3}. We also estimate the uncertainty on the median in the elemental abundance in each bin -- i.e., the uncertainty on y-axis-- using the formula $1.253\sigma/\sqrt{N}$, where $\sigma$ is the standard deviation in the distribution and $N$ is the occupation number; this formula is implemented similar to \cite{2022ApJ...928...23E}. We also tried a higher occupation number of $100$: while it reduced the number of $r_{\rm GC}$ bins, especially in the inner Galaxy regions, this did not change our final result. %On the other hand, adopting a lower occupation number of $50$ or below increased the value of the uncertainty which made it difficult to discern any obvious trend. %\footnote{We also tried creating $r_{\rm GC}$ bins of equal sizes, but this resulted in different occupation number in each bin due to the sparsity of the data across the sky.}. 

To plot the $r_{\rm GC}$ vs. [X/Fe]s distributions, we repeated the above procedure, except this time using the [X/Fe] values and using only those stars that possess \texttt{X\_FE\_FLAG}$=0$. Consequently, this removes from our analysis the elements [P/Fe], [Cu/Fe] and [Yb/Fe], as only a handful of stars possessed the corresponding \texttt{X\_FE\_FLAG}$=0$. 

Fig.~\ref{fig:Fig_3} shows that the distributions are noisy, however, many of them exhibit a noticeable transition near $r_{\rm GC}= 8\kpc$. For example, [Fe/H] shows an ascending trend from the inner Galaxy region out to $\sim8\kpc$, beyond which it shows a nearly flat trend. Some distributions show ascending trends out to $\sim8\kpc$, followed by descending trends, such as [C/Fe], [O/Fe], [Si/Fe]. Some distributions show nearly flat trends out to $\sim8\kpc$, followed by descending trends, such as [Mg/Fe], [Ca/Fe], [Ti/Fe]. Some distributions show descending trends out to $\sim8\kpc$, followed by nearly flat trends, such as [N/Fe], [Al/Fe]. Some other elements also show signs of transition near the Solar radius, but we deem these trends to be less reliable either due to large uncertainties or less number of bin points; these include [K/Fe], [Mn/Fe], [Ce/Fe]. Some distributions show some signs of transition near $r_{\rm GC}= 11\kpc$, such as [Fe/H], [C/Fe], [O/Fe], [Na/Fe], [Mg/Fe], [Cr/Fe], [TIII/Fe], [Co/Fe], [Ni/Fe].

Fig.~\ref{fig:Fig_3} is constructed mainly to highlight the trends in the metal-poor sample. In the appendix Fig.~\ref{fig:Fig_appendix3}, we compare the trends of the metal-poor sample against thick-disk and the thick-disk-to-halo samples. It clearly shows that the trends observed in the metal-poor sample are very different than those observed in the other two samples.

\section{Conclusions and Discussion}\label{sec:Discussion}

Fig.~\ref{fig:Fig_3} shows that the Milky Way's metal-poor ([Fe/H]$<-1.2$) stars that are located in the halo ($z\geq1\kpc$) display a transition in their chemical abundance profiles near the Solar radius $r_{\rm GC}=8\kpc$. We observe this trend across at least $9$ different abundances, including [Fe/H], [C/Fe], [O/Fe], [Mg/Fe], [Si/Fe], [Ca/Fe], [Ti/Fe], [N/Fe], [Al/Fe]. This trend is very unlikely to be related to radial migration, as our metal-poor sample does not contain any disk star. So, how to interpret these trends in the context of Galaxy formation?

Assuming that the Galaxy formed inside-out, $r_{\rm GC}$ provides a rough proxy for the timeline of Galaxy formation. This indicates that the observed chemical transitions near $r_{\rm GC}=8\kpc$ could be reflecting a significant change in the abundance patterns of the metal-poor stars tied to a specific epoch in the Milky Way's evolutionary history. Now, whether this change is linked to the secular chemical evolution processes or to accretion or a mix of both the processes, however, is difficult to determine from the present analysis. 

To distinguish between the secular and accretion process, one may consider an exercise of reproducing alternate versions of Fig.~\ref{fig:Fig_3} by either using only those stars associated with the known mergers or by removing these known merger stars from the data sample. This is because the general expectation is that at low metallicity intervals, especially at [Fe/H]$<-0.8$, significant fraction of stars are of accretion origin \citep[see e.g. ][]{2023MNRAS.520.5671H}. However, the reality of many of these detected mergers remains uncertain \citep[see e.g. ][]{2017A&A...604A.106J, 2025arXiv250410398T}, and their distribution functions are also not very accurately known, and perhaps not all the accretion events have yet been identified, and all of these reasons render the aforemention exercise of limited usefulness. Still, we analyze the hypothesized Gaia-Sausage/Enceladus accretion event (henceforth GSE, \citealt{2018MNRAS.478..611B, 2018Natur.563...85H}), as this stellar population has a significant presence in the inner Milky Way halo. We take the GSE sample from \cite{2024ApJ...964..104M} and this sample comprises $900$ stars. The appendix Fig.~\ref{fig:Fig_appendix4} compares the trends of the GSE sample against the metal-poor sample. The trends from the two samples are quite different for some abundances (e.g. [Fe/H], [N/Fe], [Mg/Fe], [Si/Fe]) while somewhat similar for others (e.g. [Ca/Fe], [Mn/Fe], [Ni/Fe]). %; although the similarity might be due to the large scatter and measurement uncertainties of those abundances. 
Secondly, GSE shows transition near $r_{\rm GC}=8\kpc$ for [C/Fe], [O/Fe], [Si/Fe], [Ti/Fe]. It is possible that the GSE sample is contaminated or that it comprises multiple populations. It is challenging to draw further conclusion on the trends observed in GSE. However, the overall difference observed in the GSE and metal-poor trends imply that GSE alone can not account for the metal-poor trends.

In sum, our analysis suggests that the Milky Way's metal-poor populations inside and outside the Solar radius likely originated from distinct chemical enrichment scenarios and formation processes. In the future, it will be valuable to investigate whether similar trends are present in Milky Way analogs from cosmological simulations. Additionally, it will worthwhile exploring the forthcoming spectroscopic datasets from WEAVE and 4MOST surveys, which will enable the study of such chemical trends out to larger Galactic distances.

\section*{Acknowledgements}

We thank the reviewer for helpful comments and suggestions. We thank Nicolas Martin and Chervin Laporte for their feedback on an earlier version of this manuscript.

This work has made use of data from the European Space Agency (ESA) mission {\it Gaia} (\url{https://www.cosmos.esa.int/gaia}), processed by the {\it Gaia} Data Processing and Analysis Consortium (DPAC, \url{https://www.cosmos.esa.int/web/gaia/dpac/consortium}). Funding for the DPAC has been provided by national institutions, in particular the institutions participating in the {\it Gaia} Multilateral Agreement. 

Funding for the Sloan Digital Sky Survey IV has been provided by the Alfred P. Sloan Foundation, the U.S. Department of Energy Office of Science, and the Participating Institutions. 

SDSS-IV acknowledges support and resources from the Center for High Performance Computing  at the University of Utah. The SDSS website is www.sdss4.org.

SDSS-IV is managed by the Astrophysical Research Consortium for the Participating Institutions of the SDSS Collaboration including the Brazilian Participation Group, the Carnegie Institution for Science, Carnegie Mellon University, Center for Astrophysics | Harvard \& Smithsonian, the Chilean Participation Group, the French Participation Group, Instituto de Astrof\'isica de Canarias, The Johns Hopkins  University, Kavli Institute for the Physics and Mathematics of the  Universe (IPMU) / University of Tokyo, the Korean Participation Group,  Lawrence Berkeley National Laboratory, Leibniz Institut f\"ur Astrophysik Potsdam (AIP),  Max-Planck-Institut f\"ur Astronomie (MPIA Heidelberg),  Max-Planck-Institut f\"ur Astrophysik (MPA Garching), Max-Planck-Institut f\"ur Extraterrestrische Physik (MPE), National Astronomical Observatories of China, New Mexico State University, New York University, University of Notre Dame, Observat\'ario Nacional / MCTI, The Ohio State University, Pennsylvania State University, Shanghai Astronomical Observatory, United Kingdom Participation Group, Universidad Nacional Aut\'onoma de M\'exico, University of Arizona, University of Colorado Boulder, University of Oxford, University of Portsmouth, University of Utah, University of Virginia, University of Washington, University of 
Wisconsin, Vanderbilt University, and Yale University.

\appendix

\begin{figure}[ht!]
\begin{center}
%\vspace{-0.3cm}
\hbox{
\includegraphics[width=0.5\hsize]{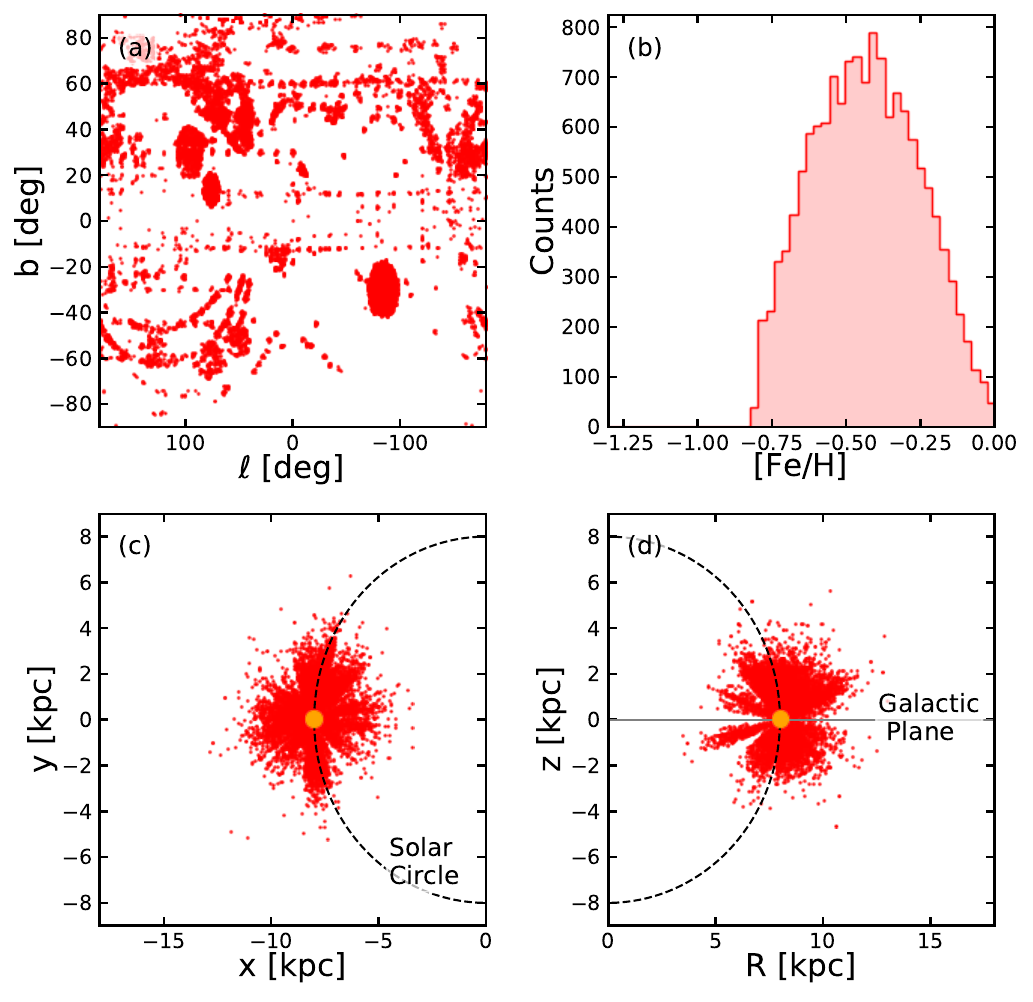}
\includegraphics[width=0.5\hsize]{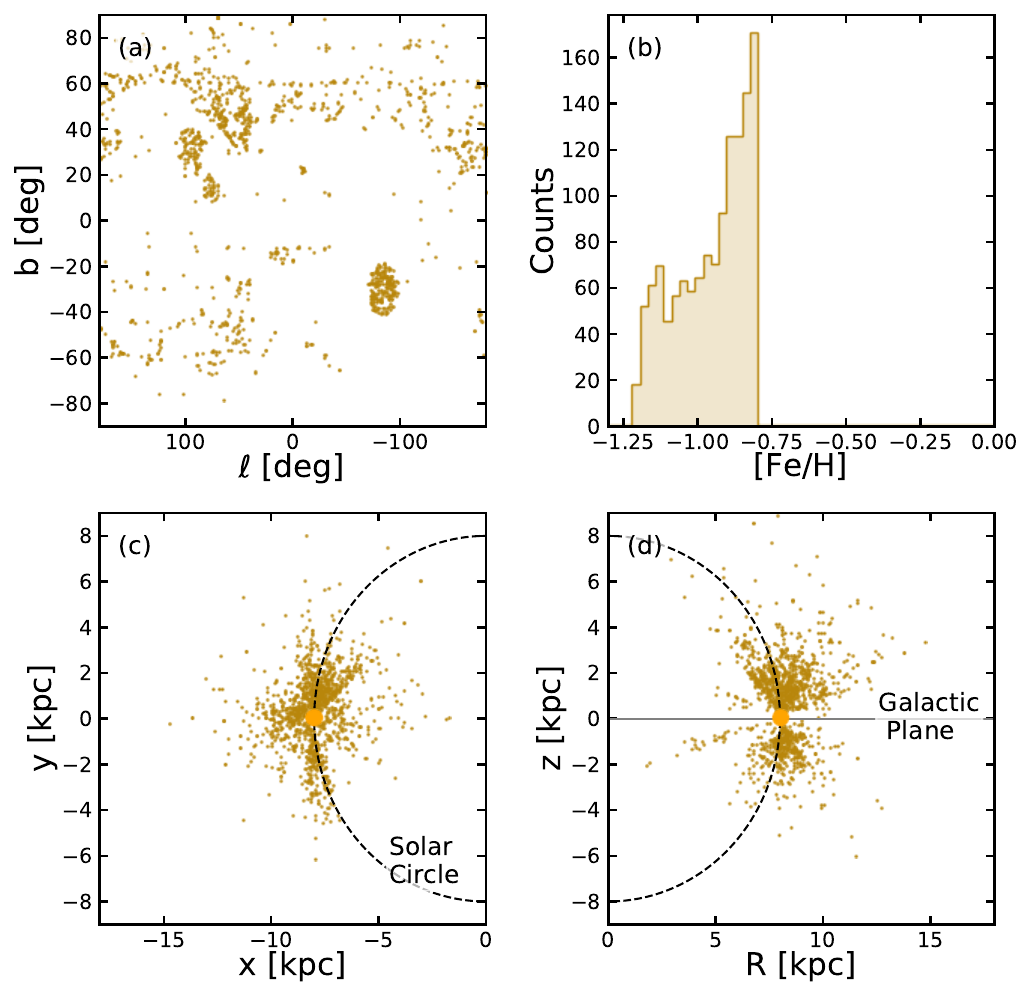}
}
\end{center}
\vspace{-0.5cm}
\caption{Same as Fig.~\ref{fig:Fig_2}, but for the thick-disk sample (top four panels) and the thick-disk-to-halo sample (bottom four panels).}
\label{fig:Fig_appendix1}
\end{figure}

\begin{figure*}
\begin{center}
%\vspace{-0.3cm}
\includegraphics[width=\hsize]{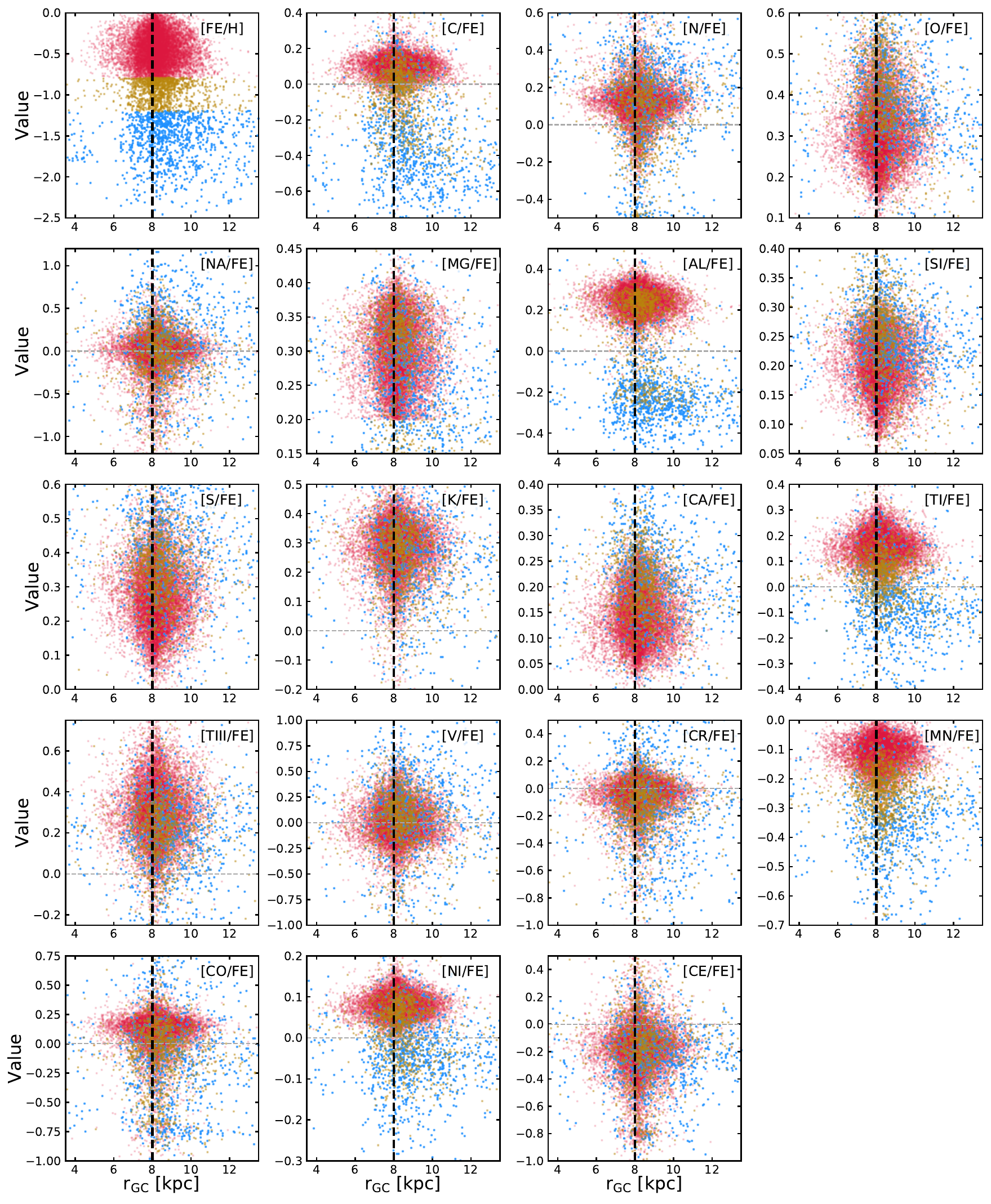}
\end{center}
\vspace{-0.5cm}
\caption{$r_{\rm GC}$ vs. [Fe/H] and $r_{\rm GC}$ vs. [X/Fe]s. Each point corresponds to a star. The distributions corresponding to the metal-poor sample are shown in blue, and those corresponding to the thick-disk sample are in red and the thick-disk-to-halo in golden. In each panel, the vertical line at $r_{\rm GC} = 8\kpc$ denotes the Solar radius. The horizontal lines (visible in some panels) are drawn where Y-axis$=0$.}
\label{fig:Fig_appendix2}
\end{figure*}

\begin{figure*}
\begin{center}
%\vspace{-0.3cm}
\includegraphics[width=\hsize]{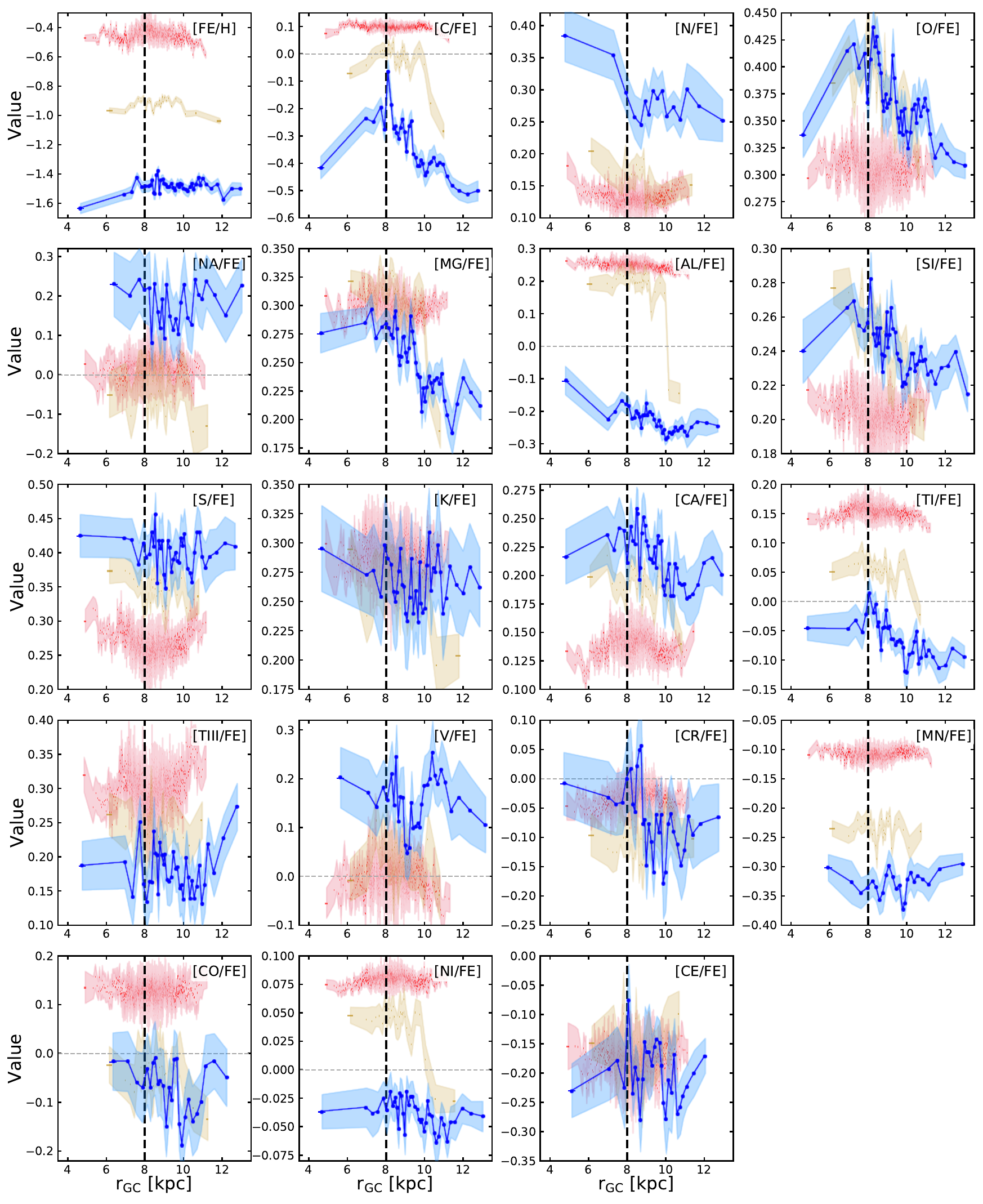}
\end{center}
\vspace{-0.5cm}
\caption{Same as Fig.~\ref{fig:Fig_3}, but now additionally showing the distributions corresponding to the thick-disk sample (in red) and the thick-disk-to-halo sample (in golden).}
\label{fig:Fig_appendix3}
\end{figure*}

\begin{figure*}
\begin{center}
%\vspace{-0.3cm}
\includegraphics[width=\hsize]{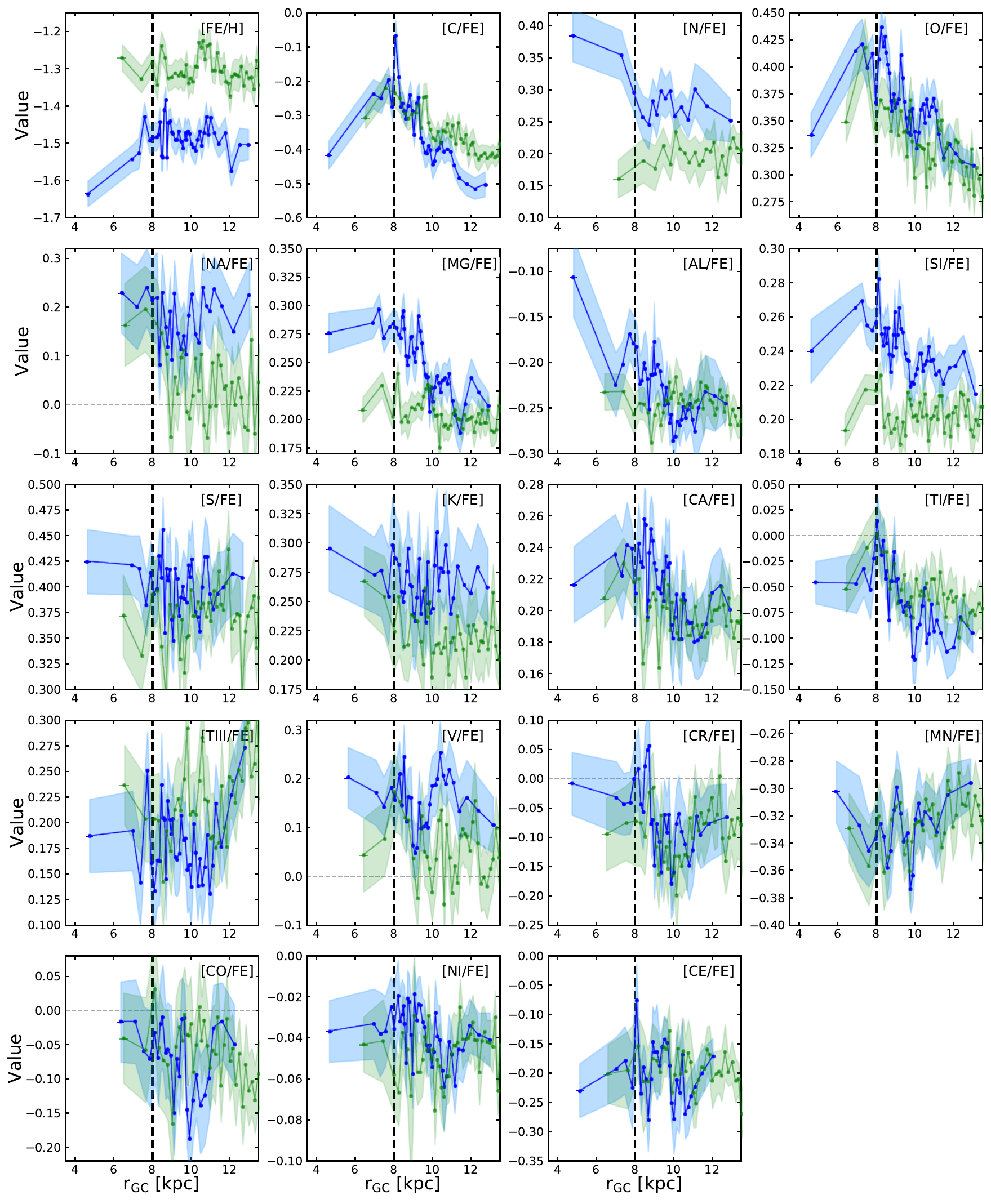}
\end{center}
\vspace{-0.5cm}
\caption{Same as Fig.~\ref{fig:Fig_3}, but now additionally showing the distributions corresponding to the GSE sample (in green).}
\label{fig:Fig_appendix4}
\end{figure*}

%
%\section{Computing stellar mass of a phase-mixed population}\label{appendix:stellar_mass}
This appendix presents Fig.~\ref{fig:Fig_appendix1}, Fig.~\ref{fig:Fig_appendix2}, Fig.~\ref{fig:Fig_appendix3} and Fig.~\ref{fig:Fig_appendix4}, as mentioned in the main text.

%

\bibliography{ref1}

\begin{thebibliography}{}
\expandafter\ifx\csname natexlab\endcsname\relax\def\natexlab#1{#1}\fi

\bibitem[{{Abdurro'uf} {et~al.}(2022){Abdurro'uf}, {Accetta}, {Aerts}, {Silva
  Aguirre}, {Ahumada}, {Ajgaonkar}, {Filiz Ak}, {Alam}, {Allende Prieto},
  {Almeida}, {Anders}, {Anderson}, {Andrews}, {Anguiano}, {Aquino-Ort{\'\i}z},
  {Arag{\'o}n-Salamanca}, {Argudo-Fern{\'a}ndez}, {Ata}, {Aubert},
  {Avila-Reese}, {Badenes}, {Barb{\'a}}, {Barger}, {Barrera-Ballesteros},
  {Beaton}, {Beers}, {Belfiore}, {Bender}, {Bernardi}, {Bershady}, {Beutler},
  {Bidin}, {Bird}, {Bizyaev}, {Blanc}, {Blanton}, {Boardman}, {Bolton},
  {Boquien}, {Borissova}, {Bovy}, {Brandt}, {Brown}, {Brownstein}, {Brusa},
  {Buchner}, {Bundy}, {Burchett}, {Bureau}, {Burgasser}, {Cabang}, {Campbell},
  {Cappellari}, {Carlberg}, {Wanderley}, {Carrera}, {Cash}, {Chen}, {Chen},
  {Cherinka}, {Chiappini}, {Choi}, {Chojnowski}, {Chung}, {Clerc}, {Cohen},
  {Comerford}, {Comparat}, {da Costa}, {Covey}, {Crane}, {Cruz-Gonzalez},
  {Culhane}, {Cunha}, {Dai}, {Damke}, {Darling}, {Davidson}, {Davies},
  {Dawson}, {De Lee}, {Diamond-Stanic}, {Cano-D{\'\i}az}, {S{\'a}nchez},
  {Donor}, {Duckworth}, {Dwelly}, {Eisenstein}, {Elsworth}, {Emsellem},
  {Eracleous}, {Escoffier}, {Fan}, {Farr}, {Feng}, {Fern{\'a}ndez-Trincado},
  {Feuillet}, {Filipp}, {Fillingham}, {Frinchaboy}, {Fromenteau}, {Galbany},
  {Garc{\'\i}a}, {Garc{\'\i}a-Hern{\'a}ndez}, {Ge}, {Geisler}, {Gelfand},
  {G{\'e}ron}, {Gibson}, {Goddy}, {Godoy-Rivera}, {Grabowski}, {Green},
  {Greener}, {Grier}, {Griffith}, {Guo}, {Guy}, {Hadjara}, {Harding},
  {Hasselquist}, {Hayes}, {Hearty}, {Hern{\'a}ndez}, {Hill}, {Hogg},
  {Holtzman}, {Horta}, {Hsieh}, {Hsu}, {Hsu}, {Huber}, {Huertas-Company},
  {Hutchinson}, {Hwang}, {Ibarra-Medel}, {Chitham}, {Ilha}, {Imig}, {Jaekle},
  {Jayasinghe}, {Ji}, {Johnson}, {Jones}, {J{\"o}nsson}, {Katkov}, {Khalatyan},
  {Kinemuchi}, {Kisku}, {Knapen}, {Kneib}, {Kollmeier}, {Kong}, {Kounkel},
  {Kreckel}, {Krishnarao}, {Lacerna}, {Lane}, {Langgin}, {Lavender}, {Law},
  {Lazarz}, {Leung}, {Leung}, {Lewis}, {Li}, {Li}, {Lian}, {Liang}, {Lin},
  {Lin}, {Lin}, {Lintott}, {Long}, {Longa-Pe{\~n}a}, {L{\'o}pez-Cob{\'a}},
  {Lu}, {Lundgren}, {Luo}, {Mackereth}, {de la Macorra}, {Mahadevan},
  {Majewski}, {Manchado}, {Mandeville}, {Maraston}, {Margalef-Bentabol},
  {Masseron}, {Masters}, {Mathur}, {McDermid}, {Mckay}, {Merloni},
  {Merrifield}, {Meszaros}, {Miglio}, {Di Mille}, {Minniti}, {Minsley}, \&
  {Monachesi}}]{2022ApJS..259...35A}
{Abdurro'uf}, {Accetta}, K., {Aerts}, C., {et~al.} 2022, \apjs, 259, 35

\bibitem[{{Arenou} {et~al.}(2018){Arenou}, {Luri}, {Babusiaux}, {Fabricius},
  {Helmi}, {Muraveva}, {Robin}, {Spoto}, {Vallenari}, {Antoja},
  {Cantat-Gaudin}, {Jordi}, {Leclerc}, {Reyl{\'e}}, {Romero-G{\'o}mez}, {Shih},
  {Soria}, {Barache}, {Bossini}, {Bragaglia}, {Breddels}, {Fabrizio},
  {Lambert}, {Marrese}, {Massari}, {Moitinho}, {Robichon}, {Ruiz-Dern},
  {Sordo}, {Veljanoski}, {Eyer}, {Jasniewicz}, {Pancino}, {Soubiran}, {Spagna},
  {Tanga}, {Turon}, \& {Zurbach}}]{2018A&A...616A..17A}
{Arenou}, F., {Luri}, X., {Babusiaux}, C., {et~al.} 2018, \aap, 616, A17

\bibitem[{{Arentsen} {et~al.}(2022){Arentsen}, {Placco}, {Lee}, {Aguado},
  {Martin}, {Starkenburg}, \& {Yoon}}]{2022MNRAS.515.4082A}
{Arentsen}, A., {Placco}, V.~M., {Lee}, Y.~S., {et~al.} 2022, \mnras, 515, 4082

\bibitem[{{Belokurov} {et~al.}(2018){Belokurov}, {Erkal}, {Evans}, {Koposov},
  \& {Deason}}]{2018MNRAS.478..611B}
{Belokurov}, V., {Erkal}, D., {Evans}, N.~W., {Koposov}, S.~E., \& {Deason},
  A.~J. 2018, \mnras, 478, 611

\bibitem[{{Belokurov} \& {Kravtsov}(2022)}]{2022MNRAS.514..689B}
{Belokurov}, V., \& {Kravtsov}, A. 2022, \mnras, 514, 689

\bibitem[{{Bressan} {et~al.}(2012){Bressan}, {Marigo}, {Girardi}, {Salasnich},
  {Dal Cero}, {Rubele}, \& {Nanni}}]{2012MNRAS.427..127B}
{Bressan}, A., {Marigo}, P., {Girardi}, L., {et~al.} 2012, \mnras, 427, 127

\bibitem[{{Buckley} {et~al.}(2024){Buckley}, {Das}, {Jofr{\'e}}, {Yates}, \&
  {Hawkins}}]{2024MNRAS.534.1985B}
{Buckley}, N., {Das}, P., {Jofr{\'e}}, P., {Yates}, R.~M., \& {Hawkins}, K.
  2024, \mnras, 534, 1985

\bibitem[{{Cerqui} {et~al.}(2023){Cerqui}, {Haywood}, {Di Matteo}, {Katz}, \&
  {Royer}}]{2023A&A...676A.108C}
{Cerqui}, V., {Haywood}, M., {Di Matteo}, P., {Katz}, D., \& {Royer}, F. 2023,
  \aap, 676, A108

\bibitem[{{Chandra} {et~al.}(2024){Chandra}, {Semenov}, {Rix}, {Conroy},
  {Bonaca}, {Naidu}, {Andrae}, {Li}, \& {Hernquist}}]{2024ApJ...972..112C}
{Chandra}, V., {Semenov}, V.~A., {Rix}, H.-W., {et~al.} 2024, \apj, 972, 112

\bibitem[{{Conroy} {et~al.}(2019){Conroy}, {Naidu}, {Zaritsky}, {Bonaca},
  {Cargile}, {Johnson}, \& {Caldwell}}]{2019ApJ...887..237C}
{Conroy}, C., {Naidu}, R.~P., {Zaritsky}, D., {et~al.} 2019, \apj, 887, 237

\bibitem[{{Da Costa} {et~al.}(2019){Da Costa}, {Bessell}, {Mackey},
  {Nordlander}, {Asplund}, {Casey}, {Frebel}, {Lind}, {Marino}, {Murphy},
  {Norris}, {Schmidt}, \& {Yong}}]{2019MNRAS.489.5900D}
{Da Costa}, G.~S., {Bessell}, M.~S., {Mackey}, A.~D., {et~al.} 2019, \mnras,
  489, 5900

\bibitem[{{Deason} \& {Belokurov}(2024)}]{2024NewAR..9901706D}
{Deason}, A.~J., \& {Belokurov}, V. 2024, \nar, 99, 101706

\bibitem[{{Di Matteo} {et~al.}(2020){Di Matteo}, {Spite}, {Haywood},
  {Bonifacio}, {G{\'o}mez}, {Spite}, \& {Caffau}}]{2020A&A...636A.115D}
{Di Matteo}, P., {Spite}, M., {Haywood}, M., {et~al.} 2020, \aap, 636, A115

\bibitem[{{Dodd} {et~al.}(2023){Dodd}, {Callingham}, {Helmi}, {Matsuno},
  {Ruiz-Lara}, {Balbinot}, \& {L{\"o}vdal}}]{2023A&A...670L...2D}
{Dodd}, E., {Callingham}, T.~M., {Helmi}, A., {et~al.} 2023, \aap, 670, L2

\bibitem[{{Eilers} {et~al.}(2022){Eilers}, {Hogg}, {Rix}, {Ness},
  {Price-Whelan}, {M{\'e}sz{\'a}ros}, \& {Nitschelm}}]{2022ApJ...928...23E}
{Eilers}, A.-C., {Hogg}, D.~W., {Rix}, H.-W., {et~al.} 2022, \apj, 928, 23

\bibitem[{{Feltzing} \& {Feuillet}(2023)}]{2023ApJ...953..143F}
{Feltzing}, S., \& {Feuillet}, D. 2023, \apj, 953, 143

\bibitem[{{Frebel} \& {Norris}(2015)}]{2015ARA&A..53..631F}
{Frebel}, A., \& {Norris}, J.~E. 2015, \araa, 53, 631

\bibitem[{{Freeman} \& {Bland-Hawthorn}(2002)}]{2002ARA&A..40..487F}
{Freeman}, K., \& {Bland-Hawthorn}, J. 2002, \araa, 40, 487

\bibitem[{{Gaia Collaboration} {et~al.}(2023){Gaia Collaboration}, {Vallenari},
  {Brown}, {Prusti}, {de Bruijne}, {Arenou}, {Babusiaux}, {Biermann},
  {Creevey}, {Ducourant}, {Evans}, {Eyer}, {Guerra}, {Hutton}, {Jordi},
  {Klioner}, {Lammers}, {Lindegren}, {Luri}, {Mignard}, {Panem}, {Pourbaix},
  {Randich}, {Sartoretti}, {Soubiran}, {Tanga}, {Walton}, {Bailer-Jones},
  {Bastian}, {Drimmel}, {Jansen}, {Katz}, {Lattanzi}, {van Leeuwen}, {Bakker},
  {Cacciari}, {Casta{\~n}eda}, {De Angeli}, {Fabricius}, {Fouesneau},
  {Fr{\'e}mat}, {Galluccio}, {Guerrier}, {Heiter}, {Masana}, {Messineo},
  {Mowlavi}, {Nicolas}, {Nienartowicz}, {Pailler}, {Panuzzo}, {Riclet}, {Roux},
  {Seabroke}, {Sordo}, {Th{\'e}venin}, {Gracia-Abril}, {Portell}, {Teyssier},
  {Altmann}, {Andrae}, {Audard}, {Bellas-Velidis}, {Benson}, {Berthier},
  {Blomme}, {Burgess}, {Busonero}, {Busso}, {C{\'a}novas}, {Carry}, {Cellino},
  {Cheek}, {Clementini}, {Damerdji}, {Davidson}, {de Teodoro}, {Nu{\~n}ez
  Campos}, {Delchambre}, {Dell'Oro}, {Esquej}, {Fern{\'a}ndez-Hern{\'a}ndez},
  {Fraile}, {Garabato}, {Garc{\'\i}a-Lario}, {Gosset}, {Haigron}, {Halbwachs},
  {Hambly}, {Harrison}, {Hern{\'a}ndez}, {Hestroffer}, {Hodgkin}, {Holl},
  {Jan{\ss}en}, {Jevardat de Fombelle}, {Jordan}, {Krone-Martins}, {Lanzafame},
  {L{\"o}ffler}, {Marchal}, {Marrese}, {Moitinho}, {Muinonen}, {Osborne},
  {Pancino}, {Pauwels}, {Recio-Blanco}, {Reyl{\'e}}, {Riello}, {Rimoldini},
  {Roegiers}, {Rybizki}, {Sarro}, {Siopis}, {Smith}, {Sozzetti}, {Utrilla},
  {van Leeuwen}, {Abbas}, {{\'A}brah{\'a}m}, {Abreu Aramburu}, {Aerts},
  {Aguado}, {Ajaj}, {Aldea-Montero}, {Altavilla}, {{\'A}lvarez}, {Alves},
  {Anders}, {Anderson}, {Anglada Varela}, {Antoja}, {Baines}, {Baker},
  {Balaguer-N{\'u}{\~n}ez}, {Balbinot}, {Balog}, {Barache}, {Barbato},
  {Barros}, {Barstow}, {Bartolom{\'e}}, {Bassilana}, {Bauchet}, {Becciani},
  {Bellazzini}, {Berihuete}, {Bernet}, {Bertone}, {Bianchi}, {Binnenfeld},
  {Blanco-Cuaresma}, {Blazere}, {Boch}, {Bombrun}, {Bossini}, {Bouquillon},
  {Bragaglia}, {Bramante}, {Breedt}, {Bressan}, {Brouillet}, {Brugaletta},
  {Bucciarelli}, {Burlacu}, {Butkevich}, {Buzzi}, {Caffau}, {Cancelliere},
  {Cantat-Gaudin}, {Carballo}, {Carlucci}, {Carnerero}, {Carrasco},
  {Casamiquela}, {Castellani}, {Castro-Ginard}, {Chaoul}, {Charlot}, {Chemin},
  {Chiaramida}, {Chiavassa}, {Chornay}, {Comoretto}, {Contursi}, {Cooper},
  {Cornez}, {Cowell}, {Crifo}, {Cropper}, {Crosta}, {Crowley}, {Dafonte},
  {Dapergolas}, {David}, {David}, {de Laverny}, {De Luise}, \& {De
  March}}]{2023A&A...674A...1G}
{Gaia Collaboration}, {Vallenari}, A., {Brown}, A.~G.~A., {et~al.} 2023, \aap,
  674, A1

\bibitem[{{GRAVITY Collaboration} {et~al.}(2019){GRAVITY Collaboration},
  {Abuter}, {Amorim}, {Baub{\"o}ck}, {Berger}, {Bonnet}, {Brandner},
  {Cl{\'e}net}, {Coud{\'e} Du Foresto}, {de Zeeuw}, {Dexter}, {Duvert},
  {Eckart}, {Eisenhauer}, {F{\"o}rster Schreiber}, {Garcia}, {Gao}, {Gendron},
  {Genzel}, {Gerhard}, {Gillessen}, {Habibi}, {Haubois}, {Henning}, {Hippler},
  {Horrobin}, {Jim{\'e}nez-Rosales}, {Jocou}, {Kervella}, {Lacour},
  {Lapeyr{\`e}re}, {Le Bouquin}, {L{\'e}na}, {Ott}, {Paumard}, {Perraut},
  {Perrin}, {Pfuhl}, {Rabien}, {Rodriguez Coira}, {Rousset}, {Scheithauer},
  {Sternberg}, {Straub}, {Straubmeier}, {Sturm}, {Tacconi}, {Vincent}, {von
  Fellenberg}, {Waisberg}, {Widmann}, {Wieprecht}, {Wiezorrek}, {Woillez}, \&
  {Yazici}}]{2019A&A...625L..10G}
{GRAVITY Collaboration}, {Abuter}, R., {Amorim}, A., {et~al.} 2019, \aap, 625,
  L10

\bibitem[{{Halle} {et~al.}(2018){Halle}, {Di Matteo}, {Haywood}, \&
  {Combes}}]{2018A&A...616A..86H}
{Halle}, A., {Di Matteo}, P., {Haywood}, M., \& {Combes}, F. 2018, \aap, 616,
  A86

\bibitem[{{Harris}(2010)}]{2010arXiv1012.3224H}
{Harris}, W.~E. 2010, arXiv e-prints, arXiv:1012.3224

\bibitem[{{Hawkins} {et~al.}(2015){Hawkins}, {Jofr{\'e}}, {Masseron}, \&
  {Gilmore}}]{2015MNRAS.453..758H}
{Hawkins}, K., {Jofr{\'e}}, P., {Masseron}, T., \& {Gilmore}, G. 2015, \mnras,
  453, 758

\bibitem[{{Haywood} {et~al.}(2018){Haywood}, {Di Matteo}, {Lehnert}, {Snaith},
  {Khoperskov}, \& {G{\'o}mez}}]{2018ApJ...863..113H}
{Haywood}, M., {Di Matteo}, P., {Lehnert}, M.~D., {et~al.} 2018, \apj, 863, 113

\bibitem[{{Haywood} {et~al.}(2024){Haywood}, {Khoperskov}, {Cerqui}, {Di
  Matteo}, {Katz}, \& {Snaith}}]{2024A&A...690A.147H}
{Haywood}, M., {Khoperskov}, S., {Cerqui}, V., {et~al.} 2024, \aap, 690, A147

\bibitem[{{Helmi}(2020)}]{2020ARA&A..58..205H}
{Helmi}, A. 2020, \araa, 58, 205

\bibitem[{{Helmi} {et~al.}(2018){Helmi}, {Babusiaux}, {Koppelman}, {Massari},
  {Veljanoski}, \& {Brown}}]{2018Natur.563...85H}
{Helmi}, A., {Babusiaux}, C., {Koppelman}, H.~H., {et~al.} 2018, \nat, 563, 85

\bibitem[{{Horta} \& {Schiavon}(2025)}]{2025MNRAS.537.3730H}
{Horta}, D., \& {Schiavon}, R.~P. 2025, \mnras, 537, 3730

\bibitem[{{Horta} {et~al.}(2023){Horta}, {Schiavon}, {Mackereth}, {Weinberg},
  {Hasselquist}, {Feuillet}, {O'Connell}, {Anguiano}, {Allende-Prieto},
  {Beaton}, {Bizyaev}, {Cunha}, {Geisler}, {Garc{\'\i}a-Hern{\'a}ndez},
  {Holtzman}, {J{\"o}nsson}, {Lane}, {Majewski}, {M{\'e}sz{\'a}ros}, {Minniti},
  {Nitschelm}, {Shetrone}, {Smith}, \& {Zasowski}}]{2023MNRAS.520.5671H}
{Horta}, D., {Schiavon}, R.~P., {Mackereth}, J.~T., {et~al.} 2023, \mnras, 520,
  5671

\bibitem[{{Hughes} {et~al.}(2022){Hughes}, {Spitler}, {Zucker}, {Nordlander},
  {Simpson}, {da Costa}, {Ting}, {Li}, {Bland-Hawthorn}, {Buder}, {Casey}, {de
  Silva}, {D'Orazi}, {Freeman}, {Hayden}, {Kos}, {Lewis}, {Lin}, {Lind},
  {Martell}, {Schlesinger}, {Sharma}, {Zwitter}, \& {GALAH
  Collaboration}}]{2022ApJ...930...47H}
{Hughes}, A. C.~N., {Spitler}, L.~R., {Zucker}, D.~B., {et~al.} 2022, \apj,
  930, 47

\bibitem[{{Jean-Baptiste} {et~al.}(2017){Jean-Baptiste}, {Di Matteo},
  {Haywood}, {G{\'o}mez}, {Montuori}, {Combes}, \&
  {Semelin}}]{2017A&A...604A.106J}
{Jean-Baptiste}, I., {Di Matteo}, P., {Haywood}, M., {et~al.} 2017, \aap, 604,
  A106

\bibitem[{{Lehmann} {et~al.}(2024){Lehmann}, {Feltzing}, {Feuillet}, \&
  {Kordopatis}}]{2024MNRAS.533..538L}
{Lehmann}, C., {Feltzing}, S., {Feuillet}, D., \& {Kordopatis}, G. 2024,
  \mnras, 533, 538

\bibitem[{{Lindegren} {et~al.}(2021){Lindegren}, {Klioner}, {Hern{\'a}ndez},
  {Bombrun}, {Ramos-Lerate}, {Steidelm{\"u}ller}, {Bastian}, {Biermann}, {de
  Torres}, {Gerlach}, {Geyer}, {Hilger}, {Hobbs}, {Lammers}, {McMillan},
  {Stephenson}, {Casta{\~n}eda}, {Davidson}, {Fabricius}, {Gracia-Abril},
  {Portell}, {Rowell}, {Teyssier}, {Torra}, {Bartolom{\'e}}, {Clotet},
  {Garralda}, {Gonz{\'a}lez-Vidal}, {Torra}, {Abbas}, {Altmann}, {Anglada
  Varela}, {Balaguer-N{\'u}{\~n}ez}, {Balog}, {Barache}, {Becciani}, {Bernet},
  {Bertone}, {Bianchi}, {Bouquillon}, {Brown}, {Bucciarelli}, {Busonero},
  {Butkevich}, {Buzzi}, {Cancelliere}, {Carlucci}, {Charlot}, {Cioni},
  {Crosta}, {Crowley}, {del Peloso}, {del Pozo}, {Drimmel}, {Esquej}, {Fienga},
  {Fraile}, {Gai}, {Garcia-Reinaldos}, {Guerra}, {Hambly}, {Hauser},
  {Jan{\ss}en}, {Jordan}, {Kostrzewa-Rutkowska}, {Lattanzi}, {Liao}, {Licata},
  {Lister}, {L{\"o}ffler}, {Marchant}, {Masip}, {Mignard}, {Mints}, {Molina},
  {Mora}, {Morbidelli}, {Murphy}, {Pagani}, {Panuzzo}, {Pe{\~n}alosa Esteller},
  {Poggio}, {Re Fiorentin}, {Riva}, {Sagrist{\`a} Sell{\'e}s}, {Sanchez
  Gimenez}, {Sarasso}, {Sciacca}, {Siddiqui}, {Smart}, {Souami}, {Spagna},
  {Steele}, {Taris}, {Utrilla}, {van Reeven}, \&
  {Vecchiato}}]{2021A&A...649A...2L}
{Lindegren}, L., {Klioner}, S.~A., {Hern{\'a}ndez}, J., {et~al.} 2021, \aap,
  649, A2

\bibitem[{{Mackereth} {et~al.}(2019){Mackereth}, {Schiavon}, {Pfeffer},
  {Hayes}, {Bovy}, {Anguiano}, {Allende Prieto}, {Hasselquist}, {Holtzman},
  {Johnson}, {Majewski}, {O'Connell}, {Shetrone}, {Tissera}, \&
  {Fern{\'a}ndez-Trincado}}]{2019MNRAS.482.3426M}
{Mackereth}, J.~T., {Schiavon}, R.~P., {Pfeffer}, J., {et~al.} 2019, \mnras,
  482, 3426

\bibitem[{{Malhan}(2022)}]{2022ApJ...930L...9M}
{Malhan}, K. 2022, \apjl, 930, L9

\bibitem[{{Malhan} \& {Rix}(2024)}]{2024ApJ...964..104M}
{Malhan}, K., \& {Rix}, H.-W. 2024, \apj, 964, 104

\bibitem[{{Matsuno} {et~al.}(2021){Matsuno}, {Hirai}, {Tarumi}, {Hotokezaka},
  {Tanaka}, \& {Helmi}}]{2021A&A...650A.110M}
{Matsuno}, T., {Hirai}, Y., {Tarumi}, Y., {et~al.} 2021, \aap, 650, A110

\bibitem[{{M{\'e}sz{\'a}ros} {et~al.}(2020){M{\'e}sz{\'a}ros}, {Masseron},
  {Garc{\'\i}a-Hern{\'a}ndez}, {Allende Prieto}, {Beers}, {Bizyaev},
  {Chojnowski}, {Cohen}, {Cunha}, {Dell'Agli}, {Ebelke},
  {Fern{\'a}ndez-Trincado}, {Frinchaboy}, {Geisler}, {Hasselquist}, {Hearty},
  {Holtzman}, {Johnson}, {Lane}, {Lacerna}, {Longa-Pe{\~n}a}, {Majewski},
  {Martell}, {Minniti}, {Nataf}, {Nidever}, {Pan}, {Schiavon}, {Shetrone},
  {Smith}, {Sobeck}, {Stringfellow}, {Szigeti}, {Tang}, {Wilson}, \&
  {Zamora}}]{2020MNRAS.492.1641M}
{M{\'e}sz{\'a}ros}, S., {Masseron}, T., {Garc{\'\i}a-Hern{\'a}ndez}, D.~A.,
  {et~al.} 2020, \mnras, 492, 1641

\bibitem[{{Monty} {et~al.}(2024){Monty}, {Belokurov}, {Sanders}, {Hansen},
  {Sakari}, {McKenzie}, {Myeong}, {Davies}, {Ardern-Arentsen}, \&
  {Massari}}]{2024MNRAS.533.2420M}
{Monty}, S., {Belokurov}, V., {Sanders}, J.~L., {et~al.} 2024, \mnras, 533,
  2420

\bibitem[{{Ratcliffe} {et~al.}(2023){Ratcliffe}, {Minchev}, {Anders},
  {Khoperskov}, {Guiglion}, {Buck}, {Cunha}, {Queiroz}, {Nitschelm},
  {Meszaros}, {Steinmetz}, {de Jong}, {Nepal}, {Lane}, \&
  {Sobeck}}]{2023MNRAS.525.2208R}
{Ratcliffe}, B., {Minchev}, I., {Anders}, F., {et~al.} 2023, \mnras, 525, 2208

\bibitem[{{Re Fiorentin} {et~al.}(2024){Re Fiorentin}, {Spagna}, {Lattanzi},
  {Cignoni}, \& {Vitali}}]{2024ApJ...977..278R}
{Re Fiorentin}, P., {Spagna}, A., {Lattanzi}, M.~G., {Cignoni}, M., \&
  {Vitali}, S. 2024, \apj, 977, 278

\bibitem[{{Rix} {et~al.}(2022){Rix}, {Chandra}, {Andrae}, {Price-Whelan},
  {Weinberg}, {Conroy}, {Fouesneau}, {Hogg}, {De Angeli}, {Naidu}, {Xiang}, \&
  {Ruz-Mieres}}]{2022ApJ...941...45R}
{Rix}, H.-W., {Chandra}, V., {Andrae}, R., {et~al.} 2022, \apj, 941, 45

\bibitem[{{Schlegel} {et~al.}(1998){Schlegel}, {Finkbeiner}, \&
  {Davis}}]{1998ApJ...500..525S}
{Schlegel}, D.~J., {Finkbeiner}, D.~P., \& {Davis}, M. 1998, \apj, 500, 525

\bibitem[{{Sellwood} \& {Binney}(2002)}]{2002MNRAS.336..785S}
{Sellwood}, J.~A., \& {Binney}, J.~J. 2002, \mnras, 336, 785

\bibitem[{{Sestito} {et~al.}(2023){Sestito}, {Venn}, {Arentsen}, {Aguado},
  {Kielty}, {Lardo}, {Martin}, {Navarro}, {Starkenburg}, {Waller}, {Carlberg},
  {Fran{\c{c}}ois}, {Gonz{\'a}lez Hern{\'a}ndez}, {Kordopatis}, {Vitali}, \&
  {Yuan}}]{2023MNRAS.518.4557S}
{Sestito}, F., {Venn}, K.~A., {Arentsen}, A., {et~al.} 2023, \mnras, 518, 4557

\bibitem[{{Sit} {et~al.}(2025){Sit}, {Weinberg}, \&
  {Griffith}}]{2025arXiv250307738S}
{Sit}, T., {Weinberg}, D.~H., \& {Griffith}, E.~J. 2025, arXiv e-prints,
  arXiv:2503.07738

\bibitem[{{Sivarani} {et~al.}(2025){Sivarani}, {Subramanian}, {Bandyopadhyay},
  {Banerjee}, {Bhattacharya}, {Choudhury}, {Ghosh}, {Hema}, {Jog}, {Hota},
  {Joshi}, {Karinkuzhi}, {Maitra}, {Malhan}, {Nayak}, {Pandey}, {Reddy},
  {Sarkar}, {Sharma}, {Singh}, {Verma}, \& {Yerra}}]{2025JApA...46...15S}
{Sivarani}, T., {Subramanian}, S., {Bandyopadhyay}, A., {et~al.} 2025, Journal
  of Astrophysics and Astronomy, 46, 15

\bibitem[{{Thomas} {et~al.}(2025){Thomas}, {Battaglia}, {Grand}, \& {Aguiar
  {\'A}lvarez}}]{2025arXiv250410398T}
{Thomas}, G.~F., {Battaglia}, G., {Grand}, R. J.~J., \& {Aguiar {\'A}lvarez},
  A. 2025, arXiv e-prints, arXiv:2504.10398

\bibitem[{{Viswanathan} {et~al.}(2024){Viswanathan}, {Bystr{\"o}m},
  {Starkenburg}, {Foppen}, {Straat}, {Montelius}, {Sestito}, {Venn},
  {Navarrete}, {Matsuno}, {Martin}, {Thomas}, {Ardern-Arentsen}, {Battaglia},
  {Fouesneau}, {Navarro}, \& {Vitali}}]{2024arXiv240817250V}
{Viswanathan}, A., {Bystr{\"o}m}, A., {Starkenburg}, E., {et~al.} 2024, arXiv
  e-prints, arXiv:2408.17250

\bibitem[{{Yuan} {et~al.}(2022){Yuan}, {Malhan}, {Sestito}, {Ibata}, {Martin},
  {Chang}, {Li}, {Caffau}, {Bonifacio}, {Bellazzini}, {Huang}, {Voggel},
  {Longeard}, {Arentsen}, {Doliva-Dolinsky}, {Navarro}, {Famaey},
  {Starkenburg}, \& {Aguado}}]{2022ApJ...930..103Y}
{Yuan}, Z., {Malhan}, K., {Sestito}, F., {et~al.} 2022, \apj, 930, 103

\bibitem[{{Zhang} {et~al.}(2021){Zhang}, {Chen}, \&
  {Zhao}}]{2021ApJ...919...52Z}
{Zhang}, H., {Chen}, Y., \& {Zhao}, G. 2021, \apj, 919, 52

\end{thebibliography}
\bibliographystyle{aasjournal}
%%%%%%%%%%%%%%%%%%%%%%%%%%%%%%%%%%%%%%%%%%%%%%%%%%

\end{document}